\documentclass[aps,pra,twocolumn,groupedaddress,showpacs]{revtex4-1}

\usepackage{graphicx}
\usepackage{amsmath,amssymb}
\usepackage{braket}
\usepackage{bm}

\begin{document}


\title{Non-equilibrium topological phase transitions in two-dimensional optical lattices}


\author{Masaya Nakagawa}
\email[]{m.nakagawa@scphys.kyoto-u.ac.jp}
\author{Norio Kawakami}
\affiliation{Department of Physics, Kyoto University, Kyoto 606-8502, Japan}


\date{\today}

\begin{abstract}
Recently, concepts of topological phases of matter are extended to non-equilibrium systems, especially periodically driven systems. In this paper, we construct an example which shows non-equilibrium topological phase transitions using ultracold fermions in optical lattices. We show that the Rabi oscillation has the possibility to induce non-equilibrium topological phases which are classified into time-reversal-invariant topological insulators for a two-orbital model of alkaline-earth-metal atoms. Furthermore we study the non-equilibrium topological phases using time-dependent Schrieffer-Wolff-type perturbation theory, and we obtain an analytical expression to describe the topological phase transitions from a high-frequency limit of external driving fields.
\end{abstract}

\pacs{67.85.-d, 37.10.Jk, 73.43.Nq}

\maketitle

\section{Introduction}
Topological properties inherent in quantum states of matter are one of the significant issues in quantum many-body physics. Starting from the studies of the quantum Hall effect, many topological phases such as topological insulators and topological superconductors \cite{Hasan,Qi} are found in solid-state systems. As well as solid-state systems, ultracold atoms in optical lattices \cite{Bloch,Dalibard,Polkovnikov} provide another fascinating stage to investigate various quantum phenomena of bosons and fermions, such as superfluidity, Mott insulators, artificially generated gauge fields, and even non-equilibrium dynamics. Some setups to realize topological phases are also proposed by using ultracold bosons and fermions \cite{Goldman,Stanescu,Gurarie,Sato,Beri,Goldman2,Mazza,Liu,Sun,Furukawa}, and experimental realization of those phases is a challenging issue in ultracold atomic physics. As a physical consequence of the topologically nontrivial nature of the quantum states, gapless edge states appear on the boundary between topologically distinct regions while the bulk energy spectrum is always gapful. These edge states yield many intriguing quantum phenomena including quantized responses, such as the quantum Hall and quantum spin Hall effect \cite{Hasan}.

Although the studies of topological phases have mostly progressed in equilibrium states, it was pointed out recently that one can investigate the topological nature of matter also in non-equilibrium quantum states, especially in periodically driven systems \cite{Oka1,Kitagawa1,Kitagawa2, Kitagawa3,Jiang,Lindner1,Lindner2,Hauke,Gomez-Leon,Rechtsman}. For example, graphene irradiated under circularly polarized light opens its band gap and shows the quantum Hall effect accompanied by gapless edge states \cite{Oka1, Kitagawa1}. Photonic quantum walks are also investigated from this viewpoint, and it is clarified that the systems can show topological phases with topologically protected zero-energy states, which is verified both theoretically and experimentally \cite{Kitagawa2,Kitagawa3}. After these works, there come new possibilities to manipulate topological properties of matter using time-dependent external fields and also to trigger topological phase transitions in non-equilibrium situations. Thus, it is a stimulating issue to understand how various time-dependent fields change the topological properties of systems, and also to explore new setups to induce non-equilibrium topological phase transitions. In addition, while there are some works on non-equilibrium topological phases, few proposals exist for realization of time-reversal-invariant topological insulators, which include quantum spin Hall insulators and $Z_2$ topological insulators, in non-equilibrium (except for Ref. \cite{Lindner1,Lindner2}). From a fundamental interest in the non-equilibrium topological quantum phenomena and for the realization of time-reversal-invariant topological insulators, it is valuable to construct a new example of non-equilibrium realization of time-reversal-invariant topological insulating phases.

In this paper, we investigate the possibilities to realize non-equilibrium topological phases using ultracold atoms. Here we demonstrate that 
the Rabi oscillation driven by external light can trigger topological phase transitions in a two-orbital optical lattice model of alkaline-earth-metal atoms. It is also shown that the Rabi oscillation preserves (effective) time-reversal symmetry, and therefore we can discuss non-equilibrium topological phases which belong to the class of time-reversal-invariant topological insulators (class AII \cite{Kitaev,Schnyder}, 
although we restrict our analyses to two-dimensional cases in this article). Furthermore, we formulate the perturbation theory for the effective Hamiltonian of periodically driven systems, and we obtain an analytical expression for the effective Hamiltonian which describes the non-equilibrium topological phase transitions.

This paper is organized as follows. In Sec. \ref{Floquet}, we summarize the formalism to describe non-equilibrium topological phases based on Floquet theory for periodically driven quantum systems. In Sec. \ref{TPT}, we show the connection between the Rabi oscillation phenomena and topological phase transitions from analytical and numerical calculations of the time-dependent Schr\"{o}dinger equation. From the (quasi-)energy spectrum, we directly confirm that the Rabi oscillation drives topological phase transitions in time-reversal-invariant two-orbital optical lattices. In Sec. \ref{ptb}, we analyze the non-equilibrium topological phase transitions from a perturbation-theoretical viewpoint to obtain more comprehensive understanding of these phenomena. In Sec. \ref{conclusion}, we summarize our results and comment on experimental realization.

\section{Floquet theory\label{Floquet}}
 In this section, we briefly review the theoretical framework to describe topological phenomena in periodically driven systems, based on Floquet theory \cite{Shirley,Sambe,Kitagawa4}. In the context of optical lattices, Floquet theory is also applied to atoms driven by periodic optical lattice shaking, which was proposed theoretically and then realized experimentally \cite{Eckardt,Struck,Lignier}. Let us consider a system driven by a time-periodic external field. The time evolution of this system is governed by a time-dependent Hamiltonian ${\cal H}(t)$, which has periodicity in time: ${\cal H}(t+T)={\cal H}(t)$ ($T$ is a period of the external field). Starting from the Schr\"{o}dinger equation $\displaystyle i \partial_t\ket{\psi (t)}={\cal H}(t)\ket{\psi (t)}$ (we set $\hbar=1$ in this paper), we can prove the following theorem (Floquet theorem): solutions of the Schr\"{o}dinger equation $\ket{\psi (t)}$ can be expressed as
\begin{gather}
\ket{\psi(t)} = e^{-i\epsilon_\alpha t}\ket{\phi_\alpha(t)}, \label{Floquet1}\\
\ket{\phi_\alpha(t+T)} = \ket{\phi_\alpha(t)}.\label{Floquet2}
\end{gather}
$\epsilon_\alpha$ is called \textit{quasi-energy}, where the index $\alpha$ labels the solution. Remarkably, the quasi-energy $\epsilon_\alpha$ and the corresponding state $\ket{\phi_\alpha (t)}$ have the following properties. Let us define an "effective Hamiltonian" ${\cal H}_\mathrm{eff}$ of this system: 
\begin{equation}
U(T,0)\equiv e^{-i{\cal H}_\mathrm{eff}T},
\label{Heff_def}
\end{equation}
where $U(t,0)={\cal T}\exp\bigl[-i\int_{0}^{t}dt'{\cal H}(t')\bigr]$ is a time-evolution operator of the system. We note that ${\cal H}_\mathrm{eff}$ is a \textit{static} Hamiltonian, not depending on time. Then, it can be shown that
\begin{gather}
{\cal H}_\mathrm{eff}\ket{\alpha}=\epsilon_\alpha\ket{\alpha}, \\
\ket{\phi_\alpha (t)}=P(t)\ket{\alpha},
\end{gather}
where $P(t)\equiv U(t,0)e^{i{\cal H}_\mathrm{eff}t}$. In other words, $\epsilon_\alpha$ is an eigenvalue of the effective Hamiltonian ${\cal H}_\mathrm{eff}$, and $\ket{\phi_\alpha (t)}$ is a "time-evolved" state of the corresponding eigenstate of ${\cal H}_\mathrm{eff}$. These facts enable us to study the time-dependent system using a static effective Hamiltonian. In a special case, if ${\cal H}_\mathrm{eff}$ has a topologically nontrivial structure in the same sense as topological insulators and topological superconductors in static systems, the system shows topological quantum phenomena. For example, the system shows topologically protected gapless states on the boundary of the system (in this formulation, "gapless" is used in the sense of the quasi-energy spectrum). Notably, these phenomena occur in genuinely time-dependent and non-equilibrium situations. When the external field causes a non-equilibrium topological phase transition, we can drive the system into topologically nontrivial states, starting from topologically trivial states.

\textit{Symmetries}.  --- Symmetries play a significant role for topological structures of Hamiltonians. To study topological quantum phenomena in periodically driven systems, we have to study symmetries of the effective Hamiltonian ${\cal H}_\mathrm{eff}$. However, from the definition Eq. (\ref{Heff_def}), we can connect symmetries of the effective Hamiltonian to those of the time-dependent Hamiltonian ${\cal H}(t)$ \cite{Lindner1,Kitagawa4}. For example, the time-reversal symmetry of the effective Hamiltonian is expressed as $Q{\cal H}_\mathrm{eff}^*Q^\dag={\cal H}_\mathrm{eff}$, where $Q$ is a unitary operator. It can be shown that the effective Hamiltonian satisfies such symmetry when ${\cal H}(t)$ satisfies
\begin{equation}
\tilde{Q}{\cal H}(-t+t_0)^*\tilde{Q}^\dag={\cal H}(t+t_0)
\label{TRS}
\end{equation}
where $Q=U^\dag (2t_0,0)\tilde{Q}$ and $t_0$ is some reference time. We can derive similar formulas for other symmetries, e.g., particle-hole symmetry, and inversion symmetry. In the following section, we use these properties to study the symmetries of the time-dependent external fields.

\section{Topological phase transitions induced by Rabi oscillation\label{TPT}}
 In this section, we show that the Rabi oscillation has the potential to change topological properties of systems, and can cause topological phase transitions in ultracold atomic systems. First, we describe the Rabi oscillation phenomena from a viewpoint of Floquet theory using a two-level system. 
We then demonstrate explicitly that the Rabi oscillation actually causes topological phase transitions in a model of two-orbital optical lattices.

\subsection{Rabi oscillation from the viewpoint of Floquet theory\label{Rabi}}
\begin{figure}
\includegraphics[width=6cm,clip]{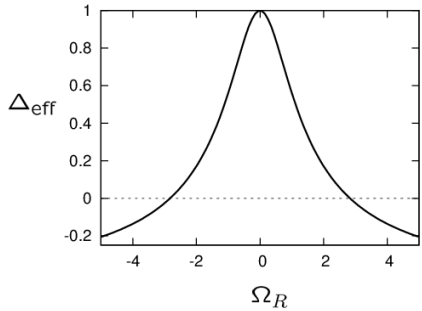}
\caption{The effective level $\Delta_\mathrm{eff}$ of the two-level system subject to the Rabi oscillation. Here we set the parameters as $\Delta=1$ (the energy unit) and $\omega=3$.\label{Deltaeff}}
\end{figure}

Let us start from a two-level system coupled with external light. The Hamiltonian is expressed as
\begin{equation}
{\cal H}(t)=
\begin{pmatrix}
\Delta & -\frac{\Omega_R}{2} e^{-i\phi} e^{-i \omega t} \\
-\frac{\Omega_R}{2} e^{i\phi} e^{i \omega t} & -\Delta
\end{pmatrix}
\label{2level}
\end{equation}
within rotating-wave approximation. Here each level has energy $+\Delta$ and $-\Delta$, and $\Omega_R$ is a Rabi frequency. $\omega$ is the frequency of the light and $\phi$ is a phase factor. This Hamiltonian is a fundamental model for Rabi oscillation, and we can solve the time-dependent Schr\"{o}dinger equation easily in a textbook manner \cite{quantum_optics}. Moreover, due to time-periodicity of this Hamiltonian ${\cal H}(t+T)={\cal H}(t)$ (where $T=2\pi/\omega$ is a period), we can apply Floquet theory to this system. From the definition Eq. (\ref{Heff_def}), we can calculate the effective Hamiltonian of this system, and the result is
\begin{equation}
{\cal H}_\mathrm{eff}=\Bigl(1-\frac{\omega}{2\epsilon_R}\Bigr)
\begin{pmatrix}
\Delta-\frac{\omega}{2} & -\frac{\Omega_R}{2} e^{-i\phi}\\
-\frac{\Omega_R}{2} e^{i\phi} & -(\Delta-\frac{\omega}{2})\\
\end{pmatrix}
\label{Rabi_Heff}
\end{equation}
where $\epsilon_R=\sqrt{(\Delta-\omega/2)^2+(\Omega_R/2)^2}$. The quasi-energy of this Hamiltonian is $\pm (\epsilon_R-\omega/2)$ (which can be interpreted as the energy of "dressed states"). We note that we can read "effective levels" $\pm \Delta_\mathrm{eff}$ from this effective Hamiltonian as 
\begin{equation}
\Delta_\mathrm{eff}=\Bigl(1-\frac{\omega}{2\epsilon_R}\Bigr)\bigl(\Delta-\frac{\omega}{2}\bigr).
\end{equation}
Naturally, $\Delta_\mathrm{eff}$ is reduced to $\Delta$ when $\Omega_R=0$. An important feature is that $\Delta_\mathrm{eff}$ reaches $\Delta-\omega/2$ as $\Omega_R\to \infty$. Then if $2\Delta<\omega$ (i.e., we apply blue-detuned external light to the system), $\Delta_\mathrm{eff}=0$ when $\Omega_R=2\sqrt{\Delta(\omega-\Delta)}$, and $\Delta_\mathrm{eff}<0$ for $\Omega_R>2\sqrt{\Delta(\omega-\Delta)}$ as shown in Fig. \ref{Deltaeff}. This is the key property of the Rabi oscillation to control topological structure in the following analyses.

\subsection{Model and symmetries}
Based on these features of an effective Hamiltonian of the Rabi oscillation phenomena, we construct a model for topological phase transitions induced by Rabi oscillation in optical lattice systems. Let us consider a band insulator of two-component fermionic atoms in an optical lattice which have two orbital degrees of freedom for each component. Two-orbital optical lattices can be implemented in several ways, for example by using alkaline-earth-metal atoms \cite{Gorshkov}, or by using higher band degrees of freedom in optical lattices \cite{Muller}. In the following, we focus on the former case. Alkaline-earth-metal atoms have the electronic ground state ($^1S_0$) and the metastable excited state ($^3P_0$), which has a long enough lifetime to neglect spontaneous emission in the typical time scale of the experiment. We can use these internal degrees of freedom as orbitals. For realizing time-reversal-invariant topological insulators, it is proper to use atoms with two spin states for each orbital, such as $^{171}$Yb, which has the nuclear spin $I=1/2$ \cite{Takasu}. These atoms are loaded in an optical lattice potential, and the two orbitals are optically coupled with each other. We assume that the laser setup to create the optical lattice has time-reversal symmetry and spatial inversion symmetry. Then the $4\times 4$ (two orbitals and two spins) Hamiltonian 
in the rotating frame of the interorbital optical coupling can be expressed as the following general form (see also similar setups considered in Ref. \cite{Beri} to realize time-reversal-invariant topological insulators in optical lattices using alkaline-earth-metal atoms):
\begin{equation}
{\cal H}_0(\bm{k})=\epsilon ({\bm k})1_4+
\begin{pmatrix}
M(\bm{k})1_2 & {\bm A}(\bm{k})\cdot \bm{\sigma} \\
{\bm A}(\bm{k})\cdot \bm{\sigma} & -M(\bm{k})1_2 \\
\end{pmatrix}.
\label{H0_band}
\end{equation}
Here $\bm{\sigma}$ are the Pauli matrices for spin and $1_n$ denotes the $n\times n$ identity matrix. $\epsilon(\bm{k})\pm M(\bm{k})$ expresses band dispersion of the upper (lower) orbital. $\bm{A}(\bm{k})$ represents the optical coupling between these orbitals. The time-reversal symmetry is expressed as $(-i\sigma_2){\cal H}_0^*(-\bm{k})(-i\sigma_2)^\dag={\cal H}_0(\bm{k})$, that is, $\epsilon(-\bm{k})=\epsilon(\bm{k})$, $M(-\bm{k})=M(\bm{k})$ and $\bm{A}(-\bm{k})=-\bm{A}(\bm{k})$. The spatial inversion symmetry is $P {\cal H}_0(-\bm{k})P^\dag={\cal H}_0(\bm{k})$, where $P=\mathrm{diag}(1,1,-1,-1)$. The Fermi energy is located between the two orbitals, such that the lower bands are completely filled and the upper bands are empty. The interorbital optical coupling $\bm{A}(\bm{k})$ is necessary to keep the system in a gapful insulator after the topological phase transition. The form ${\bm A}(\bm{k})\cdot \bm{\sigma}$ is a consequence of the fact that the polarization of the light couples with the atomic spin, and therefore this term is analogous to (but not the same as) the spin-orbit coupling in solid-state systems. 

In the following, we consider (quasi-)two-dimensional cases in which the confinement potential along the $z$ direction is sufficiently strong \cite{2D}. The detailed forms of $\epsilon(\bm{k}),M(\bm{k})$, and $\bm{A}(\bm{k})$ depend on concrete laser setups. Our strategy described below can be applied to general two-orbital band insulators in which the Hamiltonian has the form of Eq. (\ref{H0_band}), e.g., the setup in Ref. \cite{Beri}. But to make the calculation simple, we assume that one of the orbitals is $s$-like and the other is $p_x$-like (or $p_y$-like) and we start from a tight-binding description of these bands with only nearest-neighbor hopping. Then we have 
\begin{gather}
\epsilon(\bm{k})=C_1\cos \bm{k}\cdot\bm{a}_1+C_2(\cos \bm{k}\cdot\bm{a}_2+\cos \bm{k}\cdot\bm{a}_3),\\
\begin{split}
M(\bm{k})=M-&B_1(1-\cos \bm{k}\cdot\bm{a}_1)\\
&-B_2(2-\cos \bm{k}\cdot\bm{a}_2-\cos \bm{k}\cdot\bm{a}_3),
\end{split}\\
A_\mu(\bm{k})=\sum_{\nu=1}^3 A_{\mu\nu}\sin \bm{k}\cdot\bm{a}_\nu,
\end{gather}
where $\bm{a}_1=(1,0)a,\bm{a}_2=(1/2,\sqrt{3}/2)a,\bm{a}_3=(-1/2,\sqrt{3}/2)a$, and $a$ is the lattice constant. Here we set the geometry of the lattice triangular since it is rather difficult in optical lattice systems to prepare  $p_x\pm ip_y$ orbitals on square lattice geometry as in the case of semiconducting solids \cite{Hasan, Qi}. 
We set the parameter $A_{\mu\nu}$ appropriately to open a gap in the whole Brillouin zone.

Thanks to the inversion symmetry, topological properties of this system are determined by a product of parity eigenvalues at time-reversal-invariant momenta in the Brillouin zone \cite{FuKane}.  We note that the parity eigenvalues which determine the topological properties of the system are directly connected to the sign of the mass term $M(\bm{k})$ at the time-reversal-invariant momenta. This means that the system is in a topologically nontrivial phase when $0<M \lesssim 3.83B_2$ (or $3.83B_2\lesssim M<0$) in static situations. In most cases, optical lattices in such a topologically nontrivial regime are realized by strong synthetic gauge-field coupling to the atoms \cite{Beri}. However, in this article we do not assume that the static system is in the nontrivial parameter regime and try to realize the topological phase using non-equilibrium driving (Rabi oscillation), instead of using strong gauge fields. 

To the Hamiltonian of atoms in an optical lattice, we add external light as the time-dependent perturbation which induces Rabi oscillation for each atom, in the form which breaks the inversion symmetry:
\begin{equation}
{\cal H}'(t)=
\begin{pmatrix}
0 & -i\frac{\Omega_R}{2} e^{-i\phi}e^{-i\omega t} \sigma_\mu \\
i\frac{\Omega_R}{2} e^{i\phi}e^{i\omega t} \sigma_\mu & 0
\end{pmatrix},
\label{Hdash}
\end{equation}
where $\mu=1$, $2$, or $3$ indicates the direction of the polarization vector of the light, and we use the rotating-wave approximation. We note that this external perturbation is also expressed in the rotating frame of the optical coupling in Eq. (\ref{H0_band}) and therefore the frequency $\omega$ is the difference between that of the optical coupling and the external light. For simplicity, we neglect the $\bm{k}$ dependence of the Rabi frequency $\Omega_R$ and set the polarization of the external field in the direction of 
$\mu=1$. The whole system evolves in time under the Hamiltonian ${\cal H}(\bm{k},t)={\cal H}_0(\bm{k})+{\cal H}'(t)$.

We note that this external perturbation term preserves the time-reversal symmetry. From Eq. (\ref{TRS}), it can be seen that ${\cal H}'(t)$ satisfies the time-reversal symmetry for $\tilde{Q}=-i\sigma_2$ and $t_0=-\phi/\omega$. In general, ${\cal H}(t)$ and ${\cal H}_\mathrm{eff}$ have different symmetry operators, i.e., $\tilde{Q}$ and $Q$, but we can see that the symmetry operators are identical for ${\cal H}(t)$ and ${\cal H}_\mathrm{eff}$ when $\phi=0$. 

\subsection{Topological phase transitions}
Now we describe non-equilibrium topological phase transitions induced by the Rabi oscillation. In the two-level system discussed in Sec. \ref{Rabi}, the time-dependent Schr\"{o}dinger equation can be exactly solved. In the present optical-lattice system, despite neglecting interactions between atoms, we cannot solve the Schr\"{o}dinger equation easily because of the existence of the $\bm{A}(\bm{k})$ term. However, in time-reversal-invariant insulators, their topological properties are determined by some special points in the Brillouin zone, i.e., time-reversal-invariant momenta $\bm{K}$, which are equivalent to $-\bm{K}$ up to adding some multiples of reciprocal vectors \cite{FuKane}. At those points, $\bm{A}(\bm{k})$ vanishes because of time-reversal symmetry $\bm{A}(-\bm{k})=-\bm{A}(\bm{k})$. Then the Hamiltonian at time-reversal-invariant momenta is block diagonalized, and it can be seen immediately that each block is identical to the two-level systems in Eq. (\ref{2level}). 
Therefore we can solve the Schr\"{o}dinger equation at the special points, and especially we obtain an "effective mass" (corresponding to the aforementioned "effective levels") $M_\mathrm{eff}(\bm{K})$ as
\begin{equation}
M_\mathrm{eff}(\bm{K})=\Bigl(1-\frac{\omega}{2\epsilon_R}\Bigr)\bigl(M(\bm{K})-\frac{\omega}{2}\bigr),
\label{Meff}
\end{equation}
where $\epsilon_R=\sqrt{(M(\bm{K})-\omega/2)^2+(\Omega_R/2)^2}$. As mentioned before, if $2M(\bm{K})<\omega$, the effective mass at the $\bm{K}$ point can reverse its sign from positive to negative. Therefore, we can expect topological phase transitions if we appropriately choose the frequency $\omega$ and the amplitude $\Omega_R$ of the external light \cite{inversion}.

\begin{figure}
\includegraphics[width=8.5cm,clip]{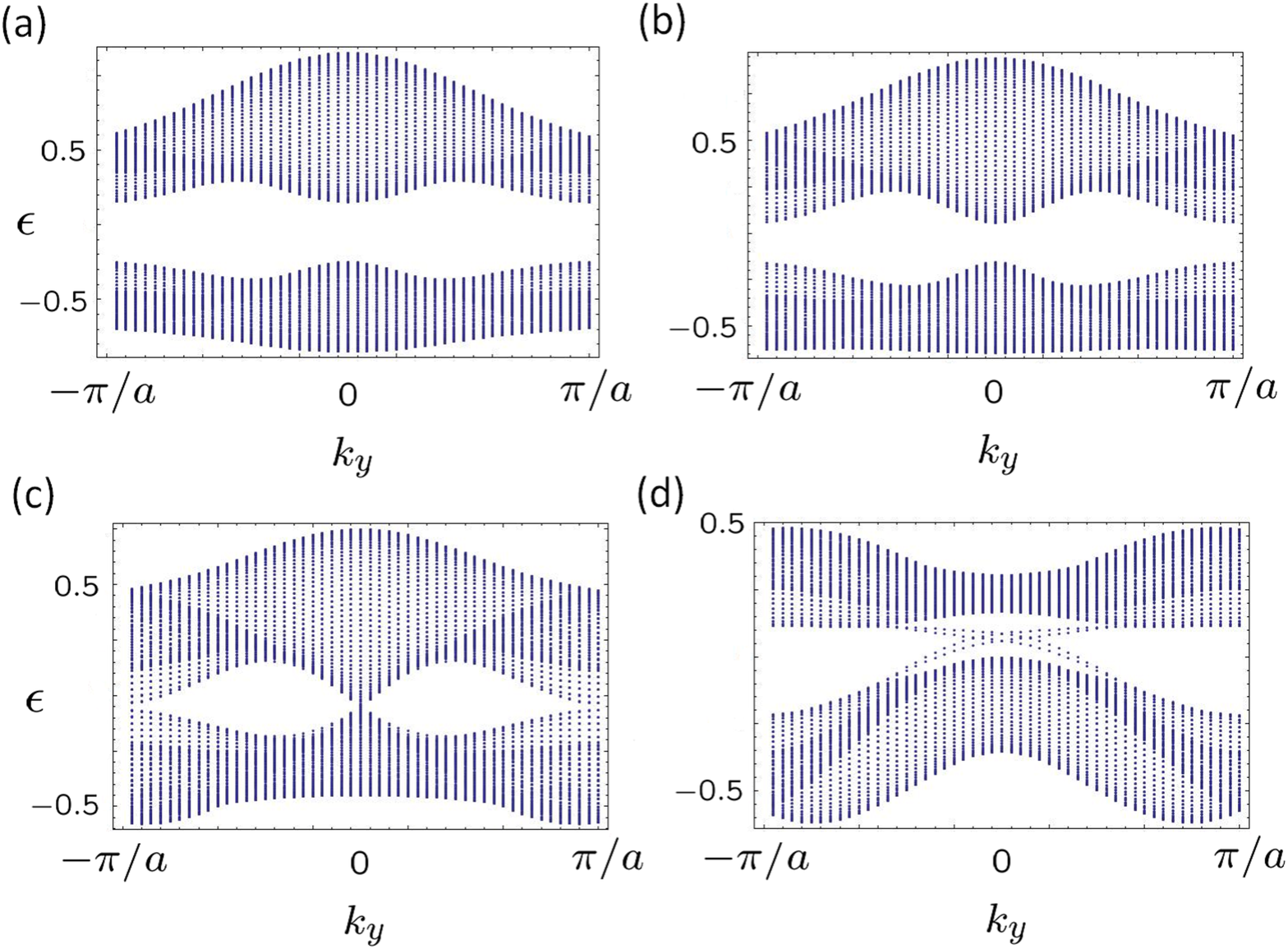}
\caption{(Color online) The quasi-energy spectrum of the effective Hamiltonian. Here we set the parameters as $C_1=C_2=0.05,M=1$ (the energy unit), $B_1=0.25,B_2=0.15,A_{11}=A_{23}=0.15,A_{13}=A_{21}=0.25,\phi=0$, and $\omega=3$. The figures show the quasi-energy at (a) $\Omega_R/2=0$, (b) $\Omega_R/2=0.5$, (c) $\Omega_R/2=0.75$, and (d) $\Omega_R/2=1.25$.\label{quasienergy}}
\end{figure}

In Fig. \ref{quasienergy}, we show the quasi-energy spectrum of this system obtained from the numerical calculation of the Schr\"{o}dinger equation under ${\cal H}(\bm{k},t)$ \cite{note}. We start from a topologically trivial band structure at $\Omega_R=0$. It can be seen apparently that the band structure changes and the band gap shrinks as we apply the external field. At $\Omega_R/2\simeq 0.75$, the band gap closes and the band structure becomes topologically nontrivial, corresponding to the sign change of the effective mass at $\bm{K}=(0,\pi/a)$. As a consequence of the topologically nontrivial band structure, it can be seen that gapless helical edge states appear between the band gap. These results mean that the (off-resonant) Rabi oscillation changes the effective band structure and actually causes the non-equilibrium topological phase transition into the time-reversal-invariant topological insulator.

\section{analysis from perturbation theory\label{ptb}}
In the previous section, we have seen the non-equilibrium topological phase transitions focusing on the Rabi oscillation at time-reversal-invariant momenta. However, it is rather unclear from the analysis how the Rabi oscillation affects the structure of the effective Hamiltonian in the whole Brillouin zone. Moreover, from a theoretical point of view, the analysis is exact at the special points in the Brillouin zone but it depends on the specific form of the time-dependent Hamiltonian, and therefore it is difficult to generalize the analysis to other dynamical phenomena. Then it should be useful to formulate another systematic method to understand the non-equilibrium topological phases induced by the Rabi oscillation in the form which can be applied to other periodically driven systems, when we investigate the non-equilibrium topological phases from a general viewpoint. In this section, we formulate the perturbation theory which can be applied to a broad class of periodically driven quantum systems, and reexamine the non-equilibrium topological phase transitions discussed in the previous section from the perturbation-theoretical viewpoint.

\subsection{Formulation of perturbation theory}
First, we construct the perturbation theory, which is useful to analyze periodically driven systems from a high-frequency limit of the external field. Roughly speaking, this perturbation theory can be regarded as a time-dependent version of the Schrieffer-Wolff transformation \cite{SW}. We start from Eq. (\ref{Floquet1}). After substituting Eq. (\ref{Floquet1}) into the Schr\"{o}dinger equation $i\partial_t\ket{\psi(t)}={\cal H}(t)\ket{\psi(t)}$, we obtain
\begin{gather}
[{\cal H}(t)-i\partial_t]\ket{\phi(t)}=\epsilon\ket{\phi(t)},\label{Floquet_eigen}\\
\ket{\phi(t+T)} = \ket{\phi(t)}.\label{Floquet_eigen2}
\end{gather}
From this equation, it can be seen that the quasi-energy and the corresponding Floquet state can be obtained from the eigenvalue problem of the operator ${\cal H}(t)-i\partial_t$ (we call it the "Floquet operator" in this paper). We note that this operator is defined on a Hilbert space $\mathbb{H}\otimes\mathbb{T}$ which includes a Hilbert space $\mathbb{T}$ of time-periodic functions ($\mathbb{H}$ is the Hilbert space of the system) \cite{Sambe}. 

The central idea of the perturbation theory is to perform a unitary transformation on the Floquet operator ${\cal H}_F\equiv {\cal H}(t)-i\partial_t$ which makes the eigenvalue problem Eq. (\ref{Floquet_eigen}) expand in terms of the external time-dependent field. We consider the unitary transformation which has a form $e^{S(t)}{\cal H}_Fe^{-S(t)}$. In this form, $S(t)$ is an anti-Hermitian operator, and must satisfy a condition $S(t+T)=S(t)$ because of the periodicity of the Floquet state [Eq. (\ref{Floquet_eigen2})]. Next, we formally expand the transformed operator in terms of $S(t)$:
\begin{widetext}
\begin{align}
e^{S(t)}{\cal H}_Fe^{-S(t)}=
&{\cal H}_0+{\cal H}'(t)-i\partial_t+[S(t),{\cal H}_0]+[S(t),{\cal H}'(t)]+[S(t),-i\partial_t]\notag\\
&+\frac{1}{2!}[S(t),[S(t),{\cal H}_0]]+\frac{1}{2!}[S(t),[S(t),{\cal H}'(t)]]+\frac{1}{2!}[S(t),[S(t),-i\partial_t]]
+\cdots.
\label{ptb_expansion}
\end{align}
\end{widetext}
Here we divide the Hamiltonian as ${\cal H}(t)={\cal H}_0+{\cal H}'(t)$, where ${\cal H}_0$ is a static part of the Hamiltonian and ${\cal H}'(t)$ is a time-periodic external field. The next step is to determine $S(t)$. We choose
\begin{equation}
[S(t),-i\partial_t]=-{\cal H}'(t)
\end{equation}
to cancel out the second term in Eq. (\ref{ptb_expansion}). Since $[S(t),-i\partial_t]=i\partial_t S(t)$, we obtain
\begin{equation}
S(t)=i\int_0^tdt'{\cal H}'(t')
\label{S}
\end{equation}
and we can readily confirm that the condition $S(t+T)=S(t)$ is satisfied if the time average of the external field over a period is zero. Then, we arrive at
\begin{align}
e^{S(t)}&{\cal H}_Fe^{-S(t)}={\cal H}_0-i\partial_t+[S(t),{\cal H}_0]+\frac{1}{2}[S(t),{\cal H}'(t)]\notag\\
&+\frac{1}{2!}[S(t),[S(t),{\cal H}_0]]+\frac{1}{3}[S(t),[S(t),{\cal H}'(t)]]+\cdots.
\label{ptb_SW}
\end{align}
From Eq. (\ref{S}), it can be seen that this expansion in $S(t)$ is justified if $r/\omega$ is small, where $r$ is the strength of the external field and $\omega=2\pi/T$ is its frequency (in the following, we assume that this condition is satisfied). Since we choose $S(t)$ to cancel out the bare external field ${\cal H}'(t)$, the transformed operator Eq. (\ref{ptb_SW}) is expressed as powers of a dimensionless parameter $r/\omega$. The operator Eq. (\ref{ptb_SW}) is still time dependent, but we note that in Eq. (\ref{ptb_SW}) the time-dependent part is smaller than the original Hamiltonian by the factor $r/\omega$ included in $S(t)$. Let us consider that the frequency is large compared to the amplitude $r$ and also large compared to the energy scale of the static part of the system, i.e., the energy scale of ${\cal H}_0$ (we denote it $\Delta$ in the following). If we assume that the contribution to the effective Hamiltonian from a time-dependent part $A$ in the Hamiltonian can be expanded in terms of a dimensionless parameter $A/\omega$, the contribution to the effective Hamiltonian from the time-dependent part of Eq. (\ref{ptb_SW}) is estimated as $\displaystyle O\bigl(\frac{r}{\omega}\cdot\frac{\Delta}{\omega}\bigr)$ and  $\displaystyle O\bigl(\bigl(\frac{r}{\omega}\bigr)^2\bigr)$. When the frequency $\omega$ is large enough, this contribution to the effective Hamiltonian is considered to be small. Therefore we can neglect the time-dependent part of Eq. (\ref{ptb_SW}) in the first order in $r/\omega$ and the zeroth order in $\Delta/\omega$ as the first approximation. This static approximation of the Floquet operator gives an approximate expression for the effective Hamiltonian. Finally, we obtain \cite{ptb_note}
\begin{align}
{\cal H}_\mathrm{eff}\simeq{\cal H}_0+&\frac{1}{T}\int_0^T dt[S(t),{\cal H}_0]+\frac{1}{2T}\int_0^T dt[S(t),{\cal H}'(t)]\notag\\
&+\frac{1}{2!T}\int_0^T dt[S(t),[S(t),{\cal H}_0]]\notag\\
&+\frac{1}{3T}\int_0^T dt[S(t),[S(t),{\cal H}'(t)]]+\cdots.
\label{ptb_formula}
\end{align}
For example, one can confirm that the expansion Eq (\ref{ptb_formula}) correctly reproduces the exact solution Eq. (\ref{Rabi_Heff}) of the Rabi oscillation in the low orders of $\Omega_R/\omega$ and $\Delta/\omega$.

We note that the perturbation formula Eq. (\ref{ptb_formula}) can be improved systematically in terms of not only the parameter $r/\omega$ but also $\Delta/\omega$. We neglect the time dependence of the transformed operator Eq. (\ref{ptb_SW}) in the zeroth order in $\Delta/\omega$ and the first order in $r/\omega$, but if we treat the time-dependent part as a \textit{new} external field and repeat the procedure of the perturbation theory, we can include the higher-order contributions of the parameter $\Delta/\omega$ and $r/\omega$. The higher-order terms in $r/\omega$ are fully included by these procedures in addition to calculating the higher-order expansion in Eq. (\ref{ptb_SW}).

\subsection{Comparison to the results of the previous section}
We now apply the perturbation theory to the system treated in the previous section. Let us express the Hamiltonian Eq. (\ref{H0_band}) as
\begin{equation}
{\cal H}_0(\bm{k})=\sum_{\mu,\nu=0}^3 d_{\mu\nu}(\bm{k})\sigma_\mu\otimes\tau_\nu.
\label{H0}
\end{equation}
Here $\sigma_\mu$ is the Pauli matrix for spins, and $\tau_\nu$ is that for orbitals (we define $\sigma_0=\tau_0=1_2$, where $1_2$ is the $2\times2$ identity matrix). Then $d_{00}(\bm{k})=\epsilon(\bm{k}),d_{03}(\bm{k})=M(\bm{k}),d_{\mu1}(\bm{k})=A_\mu(\bm{k})(\mu=1,2,3)$, and $d_{\mu\nu}(\bm{k})=0$ otherwise. We also express the external field Eq. (\ref{Hdash}) as ${\cal H}'(t)=\frac{\Omega_R}{2}(-\sin(\omega t+\phi)\sigma_1\otimes\tau_1+\cos(\omega t+\phi)\sigma_1\otimes\tau_2)$. For simplicity, we set $\phi=0$, and apply the perturbation theory to this system. For example, we obtain in the lowest order of $S(t)$,
\begin{align}
\frac{1}{T}&\int_0^T dt[S(t),{\cal H}_0]=-\frac{\Omega_R}{\omega}M(\bm{k})\sigma_1\otimes\tau_2\notag\\
&+\frac{\Omega_R}{\omega}A_2(\bm{k})\sigma_3\otimes\tau_0
-\frac{\Omega_R}{\omega}A_3(\bm{k})\sigma_2\otimes\tau_0,
\label{SH0}
\end{align}
\begin{equation}
\frac{1}{2T}\int_0^T dt[S(t),{\cal H}'(t)]=-\frac{\Omega_R^2}{4\omega}\sigma_0\otimes\tau_3.
\label{SHdash}
\end{equation}
From Eq. (\ref{SHdash}), we can notice that the contribution from $[S(t),{\cal H}'(t)]$ (in other words, the non-commutativity of the external fields in different times) gives the change of the mass term $M(\bm{k})$, which is important for the topological phase transition in the previous section. We calculate the effects of the external fields in the second order of $(\Omega_R/2)/\omega$ and in the first order of $\Delta/\omega$ by performing the expansion Eq. (\ref{ptb_formula}) in the second order of $S(t)$ and repeating the perturbative procedure once. Then, we end up with
\begin{align}
{\cal H}_\mathrm{eff}&\simeq\sum_{\mu,\nu=0}^3 d^{\mathrm{eff}}_{\mu\nu}(\bm{k})\sigma_\mu\otimes\tau_\nu,
\end{align}
where
\begin{align}
d^{\mathrm{eff}}_{00}(\bm{k})=&\epsilon(\bm{k}),\label{d00}\\
d^{\mathrm{eff}}_{03}(\bm{k})=&M(\bm{k})\Bigl(1-\frac{\Omega_R^2}{\omega^2}\Bigr)-\frac{\Omega_R^2}{4\omega},\label{d03}\\
d^{\mathrm{eff}}_{11}(\bm{k})=&A_1(\bm{k})\Bigl(1-\frac{\Omega_R^2}{4\omega^2}\Bigr),\label{d11}\\
d^{\mathrm{eff}}_{21}(\bm{k})=&A_2(\bm{k})\Bigl(1-\frac{3\Omega_R^2}{4\omega^2}\Bigr),\label{d21}\\
d^{\mathrm{eff}}_{31}(\bm{k})=&A_3(\bm{k})\Bigl(1-\frac{3\Omega_R^2}{4\omega^2}\Bigr),\label{d31}\\
d^{\mathrm{eff}}_{20}(\bm{k})=&-A_3(\bm{k})\Bigl(\frac{\Omega_R}{\omega}+\frac{2M(\bm{k})\Omega_R}{\omega^2}\Bigr),\label{d20}\\
d^{\mathrm{eff}}_{30}(\bm{k})=&A_2(\bm{k})\Bigl(\frac{\Omega_R}{\omega}+\frac{2M(\bm{k})\Omega_R}{\omega^2}\Bigr),\label{d30}\\
d^{\mathrm{eff}}_{12}(\bm{k})=&-M(\bm{k})\Bigl(\frac{\Omega_R}{\omega}+\frac{2M(\bm{k})\Omega_R}{\omega^2}\Bigr)\notag\\
&+\frac{\Omega_R^3}{4\omega^2}-\frac{2A_1^2(\bm{k})\Omega_R}{\omega^2},\label{d12}\\
d^{\mathrm{eff}}_{22}(\bm{k})=&-\frac{2A_1(\bm{k})A_2(\bm{k})\Omega_R}{\omega^2},\label{d22}\\
d^{\mathrm{eff}}_{32}(\bm{k})=&-\frac{2A_1(\bm{k})A_3(\bm{k})\Omega_R}{\omega^2}.\label{d32}
\end{align}
The terms Eqs. (\ref{d00})-(\ref{d31}) are already present in the original Hamiltonian Eq. (\ref{H0}) but modified by the external perturbation. In the special case, the $d^\mathrm{eff}_{03}(\bm{k})$ component Eq. (\ref{d03}) at time-reversal-invariant momenta $\bm{k}=\bm{K}$ indeed agrees with the expansion of the exact result Eq. (\ref{Meff}) in terms of $\Omega_R/\omega$ and $M(\bm{K})/\omega$. 
On the other hand, the terms Eqs. (\ref{d20})-(\ref{d32}) are absent in the original Hamiltonian, and they predict that the new contribution appears in the effective Hamiltonian. Particularly, Eqs. (\ref{d20}), (\ref{d30}), (\ref{d22}), and (\ref{d32}) vanish at $\bm{k}=\bm{K}$, and therefore we cannot extract these terms from the exact solution at the time-reversal-invariant momenta. 

\begin{figure}
\includegraphics[width=8cm,clip]{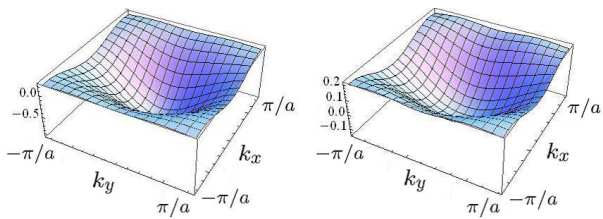}
\caption{(Color online) Plots of some components of the effective Hamiltonian. The left figure is the $\sigma_1\otimes\tau_2$ component calculated by the perturbation theory (Eq. (\ref{d12})), and the right figure is that of the numerical result for the effective Hamiltonian. The parameters of the Hamiltonian are the same as in Fig. \ref{quasienergy} (d).\label{Heff1}}
\end{figure}

\begin{figure}
\includegraphics[width=8cm,clip]{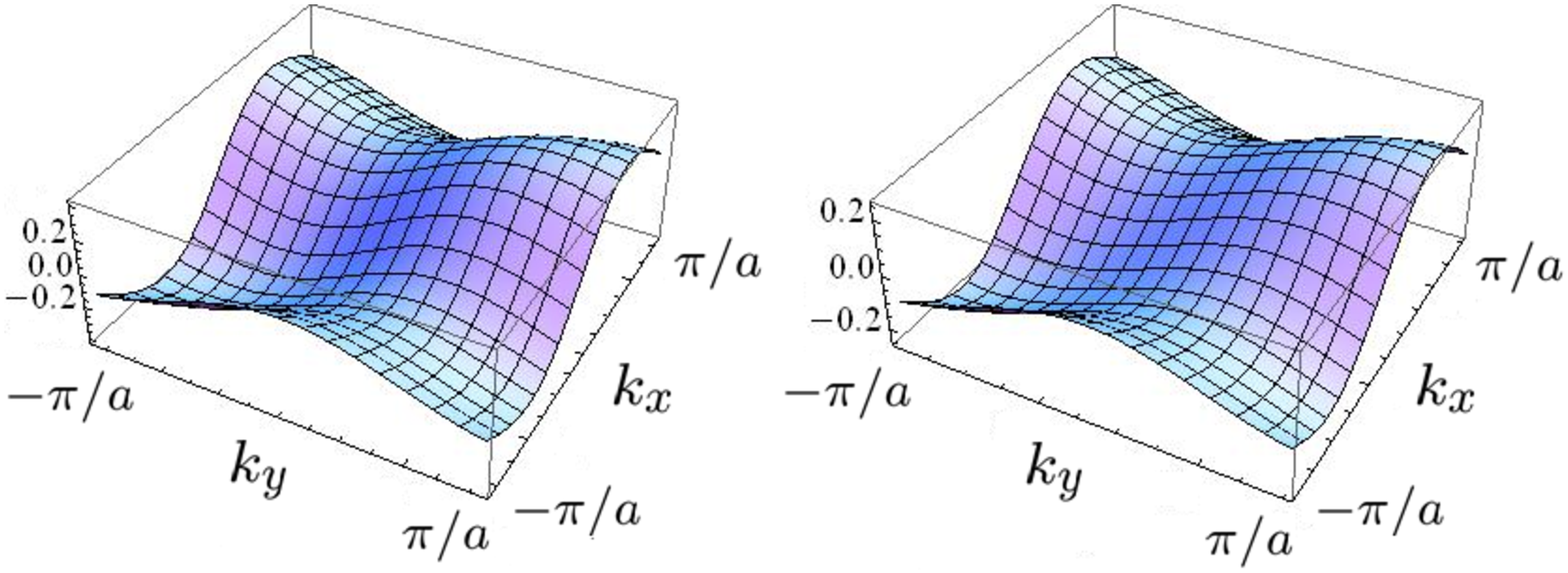}
\caption{(Color online) Plots of some components of the effective Hamiltonian. The left figure is the $\sigma_3\otimes\tau_0$ component calculated by the perturbation theory (Eq. (\ref{d30})), and the right figure is that of the numerical result for the effective Hamiltonian. The parameters of the Hamiltonian are the same as in Fig. \ref{quasienergy} (d).\label{Heff2}}
\end{figure}

We can directly check the prediction of the perturbation theory by comparing these results to the numerical calculation of the effective Hamiltonian. In Fig. \ref{Heff1} and Fig. \ref{Heff2}, we show the numerical results of the effective Hamiltonian and the corresponding components calculated from the perturbation theory. We can see clearly that the perturbative expression grasps the qualitative behavior of the effective Hamiltonian, except for quantitative difference from the higher-order contribution. The results indicate that the perturbation theory can describe the qualitative behavior of the effective Hamiltonian in the whole Brillouin zone, starting from the high-frequency limit of the external field.

\section{Conclusion\label{conclusion}}
In this paper, we have constructed a model which shows non-equilibrium topological phase transitions using ultracold atoms in optical lattices. By using the Rabi oscillation, we can control the band structure and the band gap effectively, and induce time-reversal-invariant topological insulators in dynamical situations. Furthermore, we have formulated the perturbation theory which qualitatively reproduces the behavior of the effective Hamiltonian using the time-dependent Schrieffer-Wolff-type transformation. The perturbation theory developed here can be used for a wide variety of periodically driven systems as long as the frequency of the external field is the largest energy scale in the systems, and therefore it may be useful to analyze other time-dependent phenomena in periodically driven systems.

Towards experimental realization of the non-equilibrium topological phase, we note some required points. First, to suppress the loss of atoms filled in the lower band, the driving frequency should be chosen off resonant on the other bands. Strictly speaking, of course, if we choose the frequency off resonant, the system is far from equilibrium states and it is no more in "insulating" states. Whether we can observe the edge states in such non-equilibrium distribution is one of nontrivial problems of non-equilibrium topological states, but the possibility to observe the edge states may remain if the heating of the system is not serious. The calculations for non-equilibrium properties of the topological phases which include the non-equilibrium distribution are beyond this work and left for future problems. Secondly, although the situation is complicated because of the reason mentioned above, the temperature of the atomic gas is in general required to be sufficiently lower than the band gap of the insulator. In our system, the band gap is determined by the strength of the optical coupling $\bm{A}(\bm{k})$, which depends on experimental setups but can be estimated to be of the order of the recoil energy \cite{Beri,Gerbier}.

For the analyses in this paper, we have neglected the effects of the trap potential and interactions. However, our setup to realize non-equilibrium topological phases uses local Rabi oscillation in real space; then effects of trap potentials are considered to be almost similar to those in static topological phases, which are discussed in Ref. \cite{Buchhold} for instance. This can be seen from the perturbation formula Eq. (\ref{ptb_formula}), which shows that the effects of the combination of the external field and the trap potential vanish since the local external perturbation commutes with the on-site trap potential. The effects of interactions on topological phases can be investigated with Green's-function methods \cite{Wang1,Wang2,Wang3}. As long as the interactions are not large enough to close the energy gap in the "topological Hamiltonian" \cite{Wang3}, the topological phases survive even if the atom-atom interactions exist.

\begin{acknowledgments}
We are thankful to Yoshiro Takahashi for helpful discussion. NK thanks the Japan Society for the Promotion of Science for support through its FIRST Program and KAKENHI (Grants No. 22103005 and No. 25400366).
\end{acknowledgments}

\bibliography{Floquet_top_paper_revised.bib}

\begin{thebibliography}{48}%
\makeatletter
\providecommand \@ifxundefined [1]{%
 \@ifx{#1\undefined}
}%
\providecommand \@ifnum [1]{%
 \ifnum #1\expandafter \@firstoftwo
 \else \expandafter \@secondoftwo
 \fi
}%
\providecommand \@ifx [1]{%
 \ifx #1\expandafter \@firstoftwo
 \else \expandafter \@secondoftwo
 \fi
}%
\providecommand \natexlab [1]{#1}%
\providecommand \enquote  [1]{``#1''}%
\providecommand \bibnamefont  [1]{#1}%
\providecommand \bibfnamefont [1]{#1}%
\providecommand \citenamefont [1]{#1}%
\providecommand \href@noop [0]{\@secondoftwo}%
\providecommand \href [0]{\begingroup \@sanitize@url \@href}%
\providecommand \@href[1]{\@@startlink{#1}\@@href}%
\providecommand \@@href[1]{\endgroup#1\@@endlink}%
\providecommand \@sanitize@url [0]{\catcode `\\12\catcode `\$12\catcode
  `\&12\catcode `\#12\catcode `\^12\catcode `\_12\catcode `\%12\relax}%
\providecommand \@@startlink[1]{}%
\providecommand \@@endlink[0]{}%
\providecommand \url  [0]{\begingroup\@sanitize@url \@url }%
\providecommand \@url [1]{\endgroup\@href {#1}{\urlprefix }}%
\providecommand \urlprefix  [0]{URL }%
\providecommand \Eprint [0]{\href }%
\providecommand \doibase [0]{http://dx.doi.org/}%
\providecommand \selectlanguage [0]{\@gobble}%
\providecommand \bibinfo  [0]{\@secondoftwo}%
\providecommand \bibfield  [0]{\@secondoftwo}%
\providecommand \translation [1]{[#1]}%
\providecommand \BibitemOpen [0]{}%
\providecommand \bibitemStop [0]{}%
\providecommand \bibitemNoStop [0]{.\EOS\space}%
\providecommand \EOS [0]{\spacefactor3000\relax}%
\providecommand \BibitemShut  [1]{\csname bibitem#1\endcsname}%
\let\auto@bib@innerbib\@empty
\bibitem [{\citenamefont {Hasan}\ and\ \citenamefont {Kane}(2010)}]{Hasan}%
  \BibitemOpen
  \bibfield  {author} {\bibinfo {author} {\bibfnamefont {M.~Z.}\ \bibnamefont
  {Hasan}}\ and\ \bibinfo {author} {\bibfnamefont {C.~L.}\ \bibnamefont
  {Kane}},\ }\href@noop {} {\bibfield  {journal} {\bibinfo  {journal} {Rev.
  Mod. Phys.}\ }\textbf {\bibinfo {volume} {82}},\ \bibinfo {pages} {3045}
  (\bibinfo {year} {2010})}\BibitemShut {NoStop}%
\bibitem [{\citenamefont {Qi}\ and\ \citenamefont {Zhang}(2011)}]{Qi}%
  \BibitemOpen
  \bibfield  {author} {\bibinfo {author} {\bibfnamefont {X.~L.}\ \bibnamefont
  {Qi}}\ and\ \bibinfo {author} {\bibfnamefont {S.~C.}\ \bibnamefont {Zhang}},\
  }\href@noop {} {\bibfield  {journal} {\bibinfo  {journal} {Rev. Mod. Phys.}\
  }\textbf {\bibinfo {volume} {83}},\ \bibinfo {pages} {1057} (\bibinfo {year}
  {2011})}\BibitemShut {NoStop}%
\bibitem [{\citenamefont {Bloch}\ \emph {et~al.}(2008)\citenamefont {Bloch},
  \citenamefont {Dalibard},\ and\ \citenamefont {Zwerger}}]{Bloch}%
  \BibitemOpen
  \bibfield  {author} {\bibinfo {author} {\bibfnamefont {I.}~\bibnamefont
  {Bloch}}, \bibinfo {author} {\bibfnamefont {J.}~\bibnamefont {Dalibard}}, \
  and\ \bibinfo {author} {\bibfnamefont {W.~Z.}\ \bibnamefont {Zwerger}},\
  }\href@noop {} {\bibfield  {journal} {\bibinfo  {journal} {Rev. Mod. Phys.}\
  }\textbf {\bibinfo {volume} {80}},\ \bibinfo {pages} {885} (\bibinfo {year}
  {2008})}\BibitemShut {NoStop}%
\bibitem [{\citenamefont {Dalibard}\ \emph {et~al.}(2011)\citenamefont
  {Dalibard}, \citenamefont {Gerbier}, \citenamefont {Juzeli{\={u}}nas},\ and\
  \citenamefont {{\"{O}}hberg}}]{Dalibard}%
  \BibitemOpen
  \bibfield  {author} {\bibinfo {author} {\bibfnamefont {J.}~\bibnamefont
  {Dalibard}}, \bibinfo {author} {\bibfnamefont {F.}~\bibnamefont {Gerbier}},
  \bibinfo {author} {\bibfnamefont {G.}~\bibnamefont {Juzeli{\={u}}nas}}, \
  and\ \bibinfo {author} {\bibfnamefont {P.}~\bibnamefont {{\"{O}}hberg}},\
  }\href@noop {} {\bibfield  {journal} {\bibinfo  {journal} {Rev. Mod. Phys.}\
  }\textbf {\bibinfo {volume} {83}},\ \bibinfo {pages} {1523} (\bibinfo {year}
  {2011})}\BibitemShut {NoStop}%
\bibitem [{\citenamefont {Polkovnikov}\ \emph {et~al.}(2011)\citenamefont
  {Polkovnikov}, \citenamefont {Sengupta}, \citenamefont {Silva},\ and\
  \citenamefont {Vengalattore}}]{Polkovnikov}%
  \BibitemOpen
  \bibfield  {author} {\bibinfo {author} {\bibfnamefont {A.}~\bibnamefont
  {Polkovnikov}}, \bibinfo {author} {\bibfnamefont {K.}~\bibnamefont
  {Sengupta}}, \bibinfo {author} {\bibfnamefont {A.}~\bibnamefont {Silva}}, \
  and\ \bibinfo {author} {\bibfnamefont {M.}~\bibnamefont {Vengalattore}},\
  }\href@noop {} {\bibfield  {journal} {\bibinfo  {journal} {Rev. Mod. Phys.}\
  }\textbf {\bibinfo {volume} {83}},\ \bibinfo {pages} {863} (\bibinfo {year}
  {2011})}\BibitemShut {NoStop}%
\bibitem [{\citenamefont {Goldman}\ \emph {et~al.}(2010)\citenamefont
  {Goldman}, \citenamefont {Satija}, \citenamefont {Nikolic}, \citenamefont
  {Bermudez}, \citenamefont {Martin-Delgado}, \citenamefont {Lewenstein},\ and\
  \citenamefont {Spielman}}]{Goldman}%
  \BibitemOpen
  \bibfield  {author} {\bibinfo {author} {\bibfnamefont {N.}~\bibnamefont
  {Goldman}}, \bibinfo {author} {\bibfnamefont {I.}~\bibnamefont {Satija}},
  \bibinfo {author} {\bibfnamefont {P.}~\bibnamefont {Nikolic}}, \bibinfo
  {author} {\bibfnamefont {A.}~\bibnamefont {Bermudez}}, \bibinfo {author}
  {\bibfnamefont {M.~A.}\ \bibnamefont {Martin-Delgado}}, \bibinfo {author}
  {\bibfnamefont {M.}~\bibnamefont {Lewenstein}}, \ and\ \bibinfo {author}
  {\bibfnamefont {I.~B.}\ \bibnamefont {Spielman}},\ }\href@noop {} {\bibfield
  {journal} {\bibinfo  {journal} {Phys. Rev. Lett.}\ }\textbf {\bibinfo
  {volume} {105}},\ \bibinfo {pages} {255302} (\bibinfo {year}
  {2010})}\BibitemShut {NoStop}%
\bibitem [{\citenamefont {Stanescu}\ \emph {et~al.}(2010)\citenamefont
  {Stanescu}, \citenamefont {Galitski},\ and\ \citenamefont
  {DasSarma}}]{Stanescu}%
  \BibitemOpen
  \bibfield  {author} {\bibinfo {author} {\bibfnamefont {T.~D.}\ \bibnamefont
  {Stanescu}}, \bibinfo {author} {\bibfnamefont {V.}~\bibnamefont {Galitski}},
  \ and\ \bibinfo {author} {\bibfnamefont {S.}~\bibnamefont {DasSarma}},\
  }\href@noop {} {\bibfield  {journal} {\bibinfo  {journal} {Phys. Rev. A}\
  }\textbf {\bibinfo {volume} {82}},\ \bibinfo {pages} {013608} (\bibinfo
  {year} {2010})}\BibitemShut {NoStop}%
\bibitem [{\citenamefont {Gurarie}\ and\ \citenamefont
  {Radzihovsky}(2007)}]{Gurarie}%
  \BibitemOpen
  \bibfield  {author} {\bibinfo {author} {\bibfnamefont {V.}~\bibnamefont
  {Gurarie}}\ and\ \bibinfo {author} {\bibfnamefont {L.}~\bibnamefont
  {Radzihovsky}},\ }\href@noop {} {\bibfield  {journal} {\bibinfo  {journal}
  {Ann. Phys.}\ }\textbf {\bibinfo {volume} {322}},\ \bibinfo {pages} {2}
  (\bibinfo {year} {2007})}\BibitemShut {NoStop}%
\bibitem [{\citenamefont {Sato}\ \emph {et~al.}(2010)\citenamefont {Sato},
  \citenamefont {Takahashi},\ and\ \citenamefont {Fujimoto}}]{Sato}%
  \BibitemOpen
  \bibfield  {author} {\bibinfo {author} {\bibfnamefont {M.}~\bibnamefont
  {Sato}}, \bibinfo {author} {\bibfnamefont {Y.}~\bibnamefont {Takahashi}}, \
  and\ \bibinfo {author} {\bibfnamefont {S.}~\bibnamefont {Fujimoto}},\
  }\href@noop {} {\bibfield  {journal} {\bibinfo  {journal} {Phys. Rev. B}\
  }\textbf {\bibinfo {volume} {82}},\ \bibinfo {pages} {134521} (\bibinfo
  {year} {2010})}\BibitemShut {NoStop}%
\bibitem [{\citenamefont {B{\'{e}}ri}\ and\ \citenamefont
  {Cooper}(2011)}]{Beri}%
  \BibitemOpen
  \bibfield  {author} {\bibinfo {author} {\bibfnamefont {B.}~\bibnamefont
  {B{\'{e}}ri}}\ and\ \bibinfo {author} {\bibfnamefont {N.~R.}\ \bibnamefont
  {Cooper}},\ }\href@noop {} {\bibfield  {journal} {\bibinfo  {journal} {Phys.
  Rev. Lett.}\ }\textbf {\bibinfo {volume} {107}},\ \bibinfo {pages} {145301}
  (\bibinfo {year} {2011})}\BibitemShut {NoStop}%
\bibitem [{\citenamefont {Goldman}\ \emph {et~al.}(2011)\citenamefont
  {Goldman}, \citenamefont {Urban},\ and\ \citenamefont {Bercioux}}]{Goldman2}%
  \BibitemOpen
  \bibfield  {author} {\bibinfo {author} {\bibfnamefont {N.}~\bibnamefont
  {Goldman}}, \bibinfo {author} {\bibfnamefont {D.~F.}\ \bibnamefont {Urban}},
  \ and\ \bibinfo {author} {\bibfnamefont {D.}~\bibnamefont {Bercioux}},\
  }\href@noop {} {\bibfield  {journal} {\bibinfo  {journal} {Phys. Rev. A}\
  }\textbf {\bibinfo {volume} {83}},\ \bibinfo {pages} {063601} (\bibinfo
  {year} {2011})}\BibitemShut {NoStop}%
\bibitem [{\citenamefont {Mazza}\ \emph {et~al.}(2012)\citenamefont {Mazza},
  \citenamefont {Bermudez}, \citenamefont {Goldman}, \citenamefont {Rizzi},
  \citenamefont {Martin-Delgado},\ and\ \citenamefont {Lewenstein}}]{Mazza}%
  \BibitemOpen
  \bibfield  {author} {\bibinfo {author} {\bibfnamefont {L.}~\bibnamefont
  {Mazza}}, \bibinfo {author} {\bibfnamefont {A.}~\bibnamefont {Bermudez}},
  \bibinfo {author} {\bibfnamefont {N.}~\bibnamefont {Goldman}}, \bibinfo
  {author} {\bibfnamefont {M.}~\bibnamefont {Rizzi}}, \bibinfo {author}
  {\bibfnamefont {M.~A.}\ \bibnamefont {Martin-Delgado}}, \ and\ \bibinfo
  {author} {\bibfnamefont {M.}~\bibnamefont {Lewenstein}},\ }\href@noop {}
  {\bibfield  {journal} {\bibinfo  {journal} {New. J. Phys.}\ }\textbf
  {\bibinfo {volume} {14}},\ \bibinfo {pages} {015007} (\bibinfo {year}
  {2012})}\BibitemShut {NoStop}%
\bibitem [{\citenamefont {Liu}\ and\ \citenamefont {Hu}(2012)}]{Liu}%
  \BibitemOpen
  \bibfield  {author} {\bibinfo {author} {\bibfnamefont {X.~J.}\ \bibnamefont
  {Liu}}\ and\ \bibinfo {author} {\bibfnamefont {H.}~\bibnamefont {Hu}},\
  }\href@noop {} {\bibfield  {journal} {\bibinfo  {journal} {Phys. Rev. A}\
  }\textbf {\bibinfo {volume} {85}},\ \bibinfo {pages} {033622} (\bibinfo
  {year} {2012})}\BibitemShut {NoStop}%
\bibitem [{\citenamefont {Sun}\ \emph {et~al.}(2012)\citenamefont {Sun},
  \citenamefont {Liu}, \citenamefont {Hemmerich},\ and\ \citenamefont
  {Sarma}}]{Sun}%
  \BibitemOpen
  \bibfield  {author} {\bibinfo {author} {\bibfnamefont {K.}~\bibnamefont
  {Sun}}, \bibinfo {author} {\bibfnamefont {W.~V.}\ \bibnamefont {Liu}},
  \bibinfo {author} {\bibfnamefont {A.}~\bibnamefont {Hemmerich}}, \ and\
  \bibinfo {author} {\bibfnamefont {S.~D.}\ \bibnamefont {Sarma}},\ }\href@noop
  {} {\bibfield  {journal} {\bibinfo  {journal} {Nat. Phys.}\ }\textbf
  {\bibinfo {volume} {8}},\ \bibinfo {pages} {67} (\bibinfo {year}
  {2012})}\BibitemShut {NoStop}%
\bibitem [{\citenamefont {Furukawa}\ and\ \citenamefont
  {Ueda}(2013)}]{Furukawa}%
  \BibitemOpen
  \bibfield  {author} {\bibinfo {author} {\bibfnamefont {S.}~\bibnamefont
  {Furukawa}}\ and\ \bibinfo {author} {\bibfnamefont {M.}~\bibnamefont
  {Ueda}},\ }\href@noop {} {\bibfield  {journal} {\bibinfo  {journal} {Phys.
  Rev. Lett.}\ }\textbf {\bibinfo {volume} {111}},\ \bibinfo {pages} {090401}
  (\bibinfo {year} {2013})}\BibitemShut {NoStop}%
\bibitem [{\citenamefont {Oka}\ and\ \citenamefont {Aoki}(2009)}]{Oka1}%
  \BibitemOpen
  \bibfield  {author} {\bibinfo {author} {\bibfnamefont {T.}~\bibnamefont
  {Oka}}\ and\ \bibinfo {author} {\bibfnamefont {H.}~\bibnamefont {Aoki}},\
  }\href@noop {} {\bibfield  {journal} {\bibinfo  {journal} {Phys. Rev. B}\
  }\textbf {\bibinfo {volume} {79}},\ \bibinfo {pages} {081406(R)} (\bibinfo
  {year} {2009})}\BibitemShut {NoStop}%
\bibitem [{\citenamefont {Kitagawa}\ \emph {et~al.}(2011)\citenamefont
  {Kitagawa}, \citenamefont {Oka}, \citenamefont {Brataas}, \citenamefont
  {Fu},\ and\ \citenamefont {Demler}}]{Kitagawa1}%
  \BibitemOpen
  \bibfield  {author} {\bibinfo {author} {\bibfnamefont {T.}~\bibnamefont
  {Kitagawa}}, \bibinfo {author} {\bibfnamefont {T.}~\bibnamefont {Oka}},
  \bibinfo {author} {\bibfnamefont {A.}~\bibnamefont {Brataas}}, \bibinfo
  {author} {\bibfnamefont {L.}~\bibnamefont {Fu}}, \ and\ \bibinfo {author}
  {\bibfnamefont {E.}~\bibnamefont {Demler}},\ }\href@noop {} {\bibfield
  {journal} {\bibinfo  {journal} {Phys. Rev. B}\ }\textbf {\bibinfo {volume}
  {84}},\ \bibinfo {pages} {235108} (\bibinfo {year} {2011})}\BibitemShut
  {NoStop}%
\bibitem [{\citenamefont {Kitagawa}\ \emph
  {et~al.}(2010{\natexlab{a}})\citenamefont {Kitagawa}, \citenamefont {Rudner},
  \citenamefont {Berg},\ and\ \citenamefont {Demler}}]{Kitagawa2}%
  \BibitemOpen
  \bibfield  {author} {\bibinfo {author} {\bibfnamefont {T.}~\bibnamefont
  {Kitagawa}}, \bibinfo {author} {\bibfnamefont {M.~S.}\ \bibnamefont
  {Rudner}}, \bibinfo {author} {\bibfnamefont {E.}~\bibnamefont {Berg}}, \ and\
  \bibinfo {author} {\bibfnamefont {E.}~\bibnamefont {Demler}},\ }\href@noop {}
  {\bibfield  {journal} {\bibinfo  {journal} {Phys. Rev. A}\ }\textbf {\bibinfo
  {volume} {82}},\ \bibinfo {pages} {033429} (\bibinfo {year}
  {2010}{\natexlab{a}})}\BibitemShut {NoStop}%
\bibitem [{\citenamefont {Kitagawa}\ \emph {et~al.}(2012)\citenamefont
  {Kitagawa}, \citenamefont {Broome}, \citenamefont {Fedrizzi}, \citenamefont
  {Rudner}, \citenamefont {Berg}, \citenamefont {Kassal}, \citenamefont
  {Aspuru-Guzik}, \citenamefont {Demler},\ and\ \citenamefont
  {White}}]{Kitagawa3}%
  \BibitemOpen
  \bibfield  {author} {\bibinfo {author} {\bibfnamefont {T.}~\bibnamefont
  {Kitagawa}}, \bibinfo {author} {\bibfnamefont {M.~A.}\ \bibnamefont
  {Broome}}, \bibinfo {author} {\bibfnamefont {A.}~\bibnamefont {Fedrizzi}},
  \bibinfo {author} {\bibfnamefont {M.~S.}\ \bibnamefont {Rudner}}, \bibinfo
  {author} {\bibfnamefont {E.}~\bibnamefont {Berg}}, \bibinfo {author}
  {\bibfnamefont {I.}~\bibnamefont {Kassal}}, \bibinfo {author} {\bibfnamefont
  {A.}~\bibnamefont {Aspuru-Guzik}}, \bibinfo {author} {\bibfnamefont
  {E.}~\bibnamefont {Demler}}, \ and\ \bibinfo {author} {\bibfnamefont {A.~G.}\
  \bibnamefont {White}},\ }\href@noop {} {\bibfield  {journal} {\bibinfo
  {journal} {Nat. Commun.}\ }\textbf {\bibinfo {volume} {3}},\ \bibinfo {pages}
  {882} (\bibinfo {year} {2012})}\BibitemShut {NoStop}%
\bibitem [{\citenamefont {Jiang}\ \emph {et~al.}(2011)\citenamefont {Jiang},
  \citenamefont {Kitagawa}, \citenamefont {Alicea}, \citenamefont {Akhmerov},
  \citenamefont {Pekker}, \citenamefont {Refael}, \citenamefont {Cirac},
  \citenamefont {Demler}, \citenamefont {Lukin},\ and\ \citenamefont
  {Zoller}}]{Jiang}%
  \BibitemOpen
  \bibfield  {author} {\bibinfo {author} {\bibfnamefont {L.}~\bibnamefont
  {Jiang}}, \bibinfo {author} {\bibfnamefont {T.}~\bibnamefont {Kitagawa}},
  \bibinfo {author} {\bibfnamefont {J.}~\bibnamefont {Alicea}}, \bibinfo
  {author} {\bibfnamefont {A.~R.}\ \bibnamefont {Akhmerov}}, \bibinfo {author}
  {\bibfnamefont {D.}~\bibnamefont {Pekker}}, \bibinfo {author} {\bibfnamefont
  {G.}~\bibnamefont {Refael}}, \bibinfo {author} {\bibfnamefont {J.~I.}\
  \bibnamefont {Cirac}}, \bibinfo {author} {\bibfnamefont {E.}~\bibnamefont
  {Demler}}, \bibinfo {author} {\bibfnamefont {M.~D.}\ \bibnamefont {Lukin}}, \
  and\ \bibinfo {author} {\bibfnamefont {P.}~\bibnamefont {Zoller}},\
  }\href@noop {} {\bibfield  {journal} {\bibinfo  {journal} {Phys. Rev. Lett.}\
  }\textbf {\bibinfo {volume} {106}},\ \bibinfo {pages} {220402} (\bibinfo
  {year} {2011})}\BibitemShut {NoStop}%
\bibitem [{\citenamefont {Lindner}\ \emph {et~al.}(2011)\citenamefont
  {Lindner}, \citenamefont {Refael},\ and\ \citenamefont
  {Galitski}}]{Lindner1}%
  \BibitemOpen
  \bibfield  {author} {\bibinfo {author} {\bibfnamefont {N.~H.}\ \bibnamefont
  {Lindner}}, \bibinfo {author} {\bibfnamefont {G.}~\bibnamefont {Refael}}, \
  and\ \bibinfo {author} {\bibfnamefont {V.}~\bibnamefont {Galitski}},\
  }\href@noop {} {\bibfield  {journal} {\bibinfo  {journal} {Nat. Phys.}\
  }\textbf {\bibinfo {volume} {7}},\ \bibinfo {pages} {490} (\bibinfo {year}
  {2011})}\BibitemShut {NoStop}%
\bibitem [{\citenamefont {Lindner}\ \emph {et~al.}(2013)\citenamefont
  {Lindner}, \citenamefont {Bergman}, \citenamefont {Refael},\ and\
  \citenamefont {Galitski}}]{Lindner2}%
  \BibitemOpen
  \bibfield  {author} {\bibinfo {author} {\bibfnamefont {N.~H.}\ \bibnamefont
  {Lindner}}, \bibinfo {author} {\bibfnamefont {D.~L.}\ \bibnamefont
  {Bergman}}, \bibinfo {author} {\bibfnamefont {G.}~\bibnamefont {Refael}}, \
  and\ \bibinfo {author} {\bibfnamefont {V.}~\bibnamefont {Galitski}},\
  }\href@noop {} {\bibfield  {journal} {\bibinfo  {journal} {Phys. Rev. B}\
  }\textbf {\bibinfo {volume} {87}},\ \bibinfo {pages} {235131} (\bibinfo
  {year} {2013})}\BibitemShut {NoStop}%
\bibitem [{\citenamefont {Hauke}\ \emph {et~al.}(2012)\citenamefont {Hauke},
  \citenamefont {Tieleman}, \citenamefont {Celi}, \citenamefont
  {{\"{O}}lschl{\"{a}}ger}, \citenamefont {Simonet}, \citenamefont {Struck},
  \citenamefont {Weinberg}, \citenamefont {Windpassinger}, \citenamefont
  {Sengstock}, \citenamefont {Lewenstein},\ and\ \citenamefont
  {Eckardt}}]{Hauke}%
  \BibitemOpen
  \bibfield  {author} {\bibinfo {author} {\bibfnamefont {P.}~\bibnamefont
  {Hauke}}, \bibinfo {author} {\bibfnamefont {O.}~\bibnamefont {Tieleman}},
  \bibinfo {author} {\bibfnamefont {A.}~\bibnamefont {Celi}}, \bibinfo {author}
  {\bibfnamefont {C.}~\bibnamefont {{\"{O}}lschl{\"{a}}ger}}, \bibinfo {author}
  {\bibfnamefont {J.}~\bibnamefont {Simonet}}, \bibinfo {author} {\bibfnamefont
  {J.}~\bibnamefont {Struck}}, \bibinfo {author} {\bibfnamefont
  {M.}~\bibnamefont {Weinberg}}, \bibinfo {author} {\bibfnamefont
  {P.}~\bibnamefont {Windpassinger}}, \bibinfo {author} {\bibfnamefont
  {K.}~\bibnamefont {Sengstock}}, \bibinfo {author} {\bibfnamefont
  {M.}~\bibnamefont {Lewenstein}}, \ and\ \bibinfo {author} {\bibfnamefont
  {A.}~\bibnamefont {Eckardt}},\ }\href@noop {} {\bibfield  {journal} {\bibinfo
   {journal} {Phys. Rev. Lett.}\ }\textbf {\bibinfo {volume} {109}},\ \bibinfo
  {pages} {145301} (\bibinfo {year} {2012})}\BibitemShut {NoStop}%
\bibitem [{\citenamefont {G{\'{o}}mez-Le{\'{o}}n}\ and\ \citenamefont
  {Platero}(2013)}]{Gomez-Leon}%
  \BibitemOpen
  \bibfield  {author} {\bibinfo {author} {\bibfnamefont {A.}~\bibnamefont
  {G{\'{o}}mez-Le{\'{o}}n}}\ and\ \bibinfo {author} {\bibfnamefont
  {G.}~\bibnamefont {Platero}},\ }\href@noop {} {\bibfield  {journal} {\bibinfo
   {journal} {Phys. Rev. Lett.}\ }\textbf {\bibinfo {volume} {110}},\ \bibinfo
  {pages} {200403} (\bibinfo {year} {2013})}\BibitemShut {NoStop}%
\bibitem [{\citenamefont {Rechtsman}\ \emph {et~al.}(2013)\citenamefont
  {Rechtsman}, \citenamefont {Zeuner}, \citenamefont {Plotnik}, \citenamefont
  {Lumer}, \citenamefont {Podolsky}, \citenamefont {Dreisow}, \citenamefont
  {Nolte}, \citenamefont {Segev},\ and\ \citenamefont {Szameit}}]{Rechtsman}%
  \BibitemOpen
  \bibfield  {author} {\bibinfo {author} {\bibfnamefont {M.~C.}\ \bibnamefont
  {Rechtsman}}, \bibinfo {author} {\bibfnamefont {J.~M.}\ \bibnamefont
  {Zeuner}}, \bibinfo {author} {\bibfnamefont {Y.}~\bibnamefont {Plotnik}},
  \bibinfo {author} {\bibfnamefont {Y.}~\bibnamefont {Lumer}}, \bibinfo
  {author} {\bibfnamefont {D.}~\bibnamefont {Podolsky}}, \bibinfo {author}
  {\bibfnamefont {F.}~\bibnamefont {Dreisow}}, \bibinfo {author} {\bibfnamefont
  {S.}~\bibnamefont {Nolte}}, \bibinfo {author} {\bibfnamefont
  {M.}~\bibnamefont {Segev}}, \ and\ \bibinfo {author} {\bibfnamefont
  {A.}~\bibnamefont {Szameit}},\ }\href@noop {} {\bibfield  {journal} {\bibinfo
   {journal} {Nature}\ }\textbf {\bibinfo {volume} {496}},\ \bibinfo {pages}
  {196} (\bibinfo {year} {2013})}\BibitemShut {NoStop}%
\bibitem [{\citenamefont {Kitaev}(2009)}]{Kitaev}%
  \BibitemOpen
  \bibfield  {author} {\bibinfo {author} {\bibfnamefont {A.}~\bibnamefont
  {Kitaev}},\ }\href@noop {} {\bibfield  {journal} {\bibinfo  {journal} {AIP.
  Conf. Proc.}\ }\textbf {\bibinfo {volume} {1134}},\ \bibinfo {pages} {22}
  (\bibinfo {year} {2009})}\BibitemShut {NoStop}%
\bibitem [{\citenamefont {Schnyder}\ \emph {et~al.}(2008)\citenamefont
  {Schnyder}, \citenamefont {Ryu}, \citenamefont {Furusaki},\ and\
  \citenamefont {Ludwig}}]{Schnyder}%
  \BibitemOpen
  \bibfield  {author} {\bibinfo {author} {\bibfnamefont {A.~P.}\ \bibnamefont
  {Schnyder}}, \bibinfo {author} {\bibfnamefont {S.}~\bibnamefont {Ryu}},
  \bibinfo {author} {\bibfnamefont {A.}~\bibnamefont {Furusaki}}, \ and\
  \bibinfo {author} {\bibfnamefont {A.~W.~W.}\ \bibnamefont {Ludwig}},\
  }\href@noop {} {\bibfield  {journal} {\bibinfo  {journal} {Phys. Rev. B}\
  }\textbf {\bibinfo {volume} {78}},\ \bibinfo {pages} {195125} (\bibinfo
  {year} {2008})}\BibitemShut {NoStop}%
\bibitem [{\citenamefont {Shirley}(1965)}]{Shirley}%
  \BibitemOpen
  \bibfield  {author} {\bibinfo {author} {\bibfnamefont {J.~H.}\ \bibnamefont
  {Shirley}},\ }\href@noop {} {\bibfield  {journal} {\bibinfo  {journal} {Phys.
  Rev.}\ }\textbf {\bibinfo {volume} {138}},\ \bibinfo {pages} {B979} (\bibinfo
  {year} {1965})}\BibitemShut {NoStop}%
\bibitem [{\citenamefont {Sambe}(1973)}]{Sambe}%
  \BibitemOpen
  \bibfield  {author} {\bibinfo {author} {\bibfnamefont {H.}~\bibnamefont
  {Sambe}},\ }\href@noop {} {\bibfield  {journal} {\bibinfo  {journal} {Phys.
  Rev. A}\ }\textbf {\bibinfo {volume} {7}},\ \bibinfo {pages} {2203} (\bibinfo
  {year} {1973})}\BibitemShut {NoStop}%
\bibitem [{\citenamefont {Kitagawa}\ \emph
  {et~al.}(2010{\natexlab{b}})\citenamefont {Kitagawa}, \citenamefont {Berg},
  \citenamefont {Rudner},\ and\ \citenamefont {Demler}}]{Kitagawa4}%
  \BibitemOpen
  \bibfield  {author} {\bibinfo {author} {\bibfnamefont {T.}~\bibnamefont
  {Kitagawa}}, \bibinfo {author} {\bibfnamefont {E.}~\bibnamefont {Berg}},
  \bibinfo {author} {\bibfnamefont {M.}~\bibnamefont {Rudner}}, \ and\ \bibinfo
  {author} {\bibfnamefont {E.}~\bibnamefont {Demler}},\ }\href@noop {}
  {\bibfield  {journal} {\bibinfo  {journal} {Phys. Rev. B}\ }\textbf {\bibinfo
  {volume} {82}},\ \bibinfo {pages} {235114} (\bibinfo {year}
  {2010}{\natexlab{b}})}\BibitemShut {NoStop}%
\bibitem [{\citenamefont {Eckardt}\ \emph {et~al.}(2005)\citenamefont
  {Eckardt}, \citenamefont {Weiss},\ and\ \citenamefont {Holthaus}}]{Eckardt}%
  \BibitemOpen
  \bibfield  {author} {\bibinfo {author} {\bibfnamefont {A.}~\bibnamefont
  {Eckardt}}, \bibinfo {author} {\bibfnamefont {C.}~\bibnamefont {Weiss}}, \
  and\ \bibinfo {author} {\bibfnamefont {M.}~\bibnamefont {Holthaus}},\
  }\href@noop {} {\bibfield  {journal} {\bibinfo  {journal} {Phys. Rev. Lett.}\
  }\textbf {\bibinfo {volume} {95}},\ \bibinfo {pages} {260404} (\bibinfo
  {year} {2005})}\BibitemShut {NoStop}%
\bibitem [{\citenamefont {Struck}\ \emph {et~al.}(2011)\citenamefont {Struck},
  \citenamefont {{\"{O}}lschl{\"{a}}ger}, \citenamefont {Targat}, \citenamefont
  {Soltan-Panahi}, \citenamefont {Eckardt}, \citenamefont {Lewenstein},
  \citenamefont {Windpassinger},\ and\ \citenamefont {Sengstock}}]{Struck}%
  \BibitemOpen
  \bibfield  {author} {\bibinfo {author} {\bibfnamefont {J.}~\bibnamefont
  {Struck}}, \bibinfo {author} {\bibfnamefont {C.}~\bibnamefont
  {{\"{O}}lschl{\"{a}}ger}}, \bibinfo {author} {\bibfnamefont {R.~L.}\
  \bibnamefont {Targat}}, \bibinfo {author} {\bibfnamefont {P.}~\bibnamefont
  {Soltan-Panahi}}, \bibinfo {author} {\bibfnamefont {A.}~\bibnamefont
  {Eckardt}}, \bibinfo {author} {\bibfnamefont {M.}~\bibnamefont {Lewenstein}},
  \bibinfo {author} {\bibfnamefont {P.}~\bibnamefont {Windpassinger}}, \ and\
  \bibinfo {author} {\bibfnamefont {K.}~\bibnamefont {Sengstock}},\ }\href@noop
  {} {\bibfield  {journal} {\bibinfo  {journal} {Science}\ }\textbf {\bibinfo
  {volume} {333}},\ \bibinfo {pages} {996} (\bibinfo {year}
  {2011})}\BibitemShut {NoStop}%
\bibitem [{\citenamefont {Lignier}\ \emph {et~al.}(2007)\citenamefont
  {Lignier}, \citenamefont {Sias}, \citenamefont {Ciampini}, \citenamefont
  {Singh}, \citenamefont {Zenesini}, \citenamefont {Morsch},\ and\
  \citenamefont {Arimondo}}]{Lignier}%
  \BibitemOpen
  \bibfield  {author} {\bibinfo {author} {\bibfnamefont {H.}~\bibnamefont
  {Lignier}}, \bibinfo {author} {\bibfnamefont {C.}~\bibnamefont {Sias}},
  \bibinfo {author} {\bibfnamefont {D.}~\bibnamefont {Ciampini}}, \bibinfo
  {author} {\bibfnamefont {Y.}~\bibnamefont {Singh}}, \bibinfo {author}
  {\bibfnamefont {A.}~\bibnamefont {Zenesini}}, \bibinfo {author}
  {\bibfnamefont {O.}~\bibnamefont {Morsch}}, \ and\ \bibinfo {author}
  {\bibfnamefont {E.}~\bibnamefont {Arimondo}},\ }\href@noop {} {\bibfield
  {journal} {\bibinfo  {journal} {Phys. Rev. Lett.}\ }\textbf {\bibinfo
  {volume} {99}},\ \bibinfo {pages} {220403} (\bibinfo {year}
  {2007})}\BibitemShut {NoStop}%
\bibitem [{\citenamefont {Fox}(2006)}]{quantum_optics}%
  \BibitemOpen
  \bibfield  {author} {\bibinfo {author} {\bibfnamefont {M.}~\bibnamefont
  {Fox}},\ }\href@noop {} {\emph {\bibinfo {title} {Quantum Optics: An
  Introduction}}}\ (\bibinfo  {publisher} {Oxford University Press},\ \bibinfo
  {address} {Oxford},\ \bibinfo {year} {2006})\BibitemShut {NoStop}%
\bibitem [{\citenamefont {Gorshkov}\ \emph {et~al.}(2010)\citenamefont
  {Gorshkov}, \citenamefont {Hermele}, \citenamefont {Gurarie}, \citenamefont
  {Xu}, \citenamefont {Julienne}, \citenamefont {Ye}, \citenamefont {Zoller},\
  and\ \citenamefont {Demler}}]{Gorshkov}%
  \BibitemOpen
  \bibfield  {author} {\bibinfo {author} {\bibfnamefont {A.~V.}\ \bibnamefont
  {Gorshkov}}, \bibinfo {author} {\bibfnamefont {M.}~\bibnamefont {Hermele}},
  \bibinfo {author} {\bibfnamefont {V.}~\bibnamefont {Gurarie}}, \bibinfo
  {author} {\bibfnamefont {C.}~\bibnamefont {Xu}}, \bibinfo {author}
  {\bibfnamefont {P.~S.}\ \bibnamefont {Julienne}}, \bibinfo {author}
  {\bibfnamefont {J.}~\bibnamefont {Ye}}, \bibinfo {author} {\bibfnamefont
  {P.}~\bibnamefont {Zoller}}, \ and\ \bibinfo {author} {\bibfnamefont
  {E.}~\bibnamefont {Demler}},\ }\href@noop {} {\bibfield  {journal} {\bibinfo
  {journal} {Nat. Phys.}\ }\textbf {\bibinfo {volume} {6}},\ \bibinfo {pages}
  {289} (\bibinfo {year} {2010})}\BibitemShut {NoStop}%
\bibitem [{\citenamefont {M{\"{u}}ller}\ \emph {et~al.}(2007)\citenamefont
  {M{\"{u}}ller}, \citenamefont {F{\"{o}}lling}, \citenamefont {Widera},\ and\
  \citenamefont {Bloch}}]{Muller}%
  \BibitemOpen
  \bibfield  {author} {\bibinfo {author} {\bibfnamefont {T.}~\bibnamefont
  {M{\"{u}}ller}}, \bibinfo {author} {\bibfnamefont {S.}~\bibnamefont
  {F{\"{o}}lling}}, \bibinfo {author} {\bibfnamefont {A.}~\bibnamefont
  {Widera}}, \ and\ \bibinfo {author} {\bibfnamefont {I.}~\bibnamefont
  {Bloch}},\ }\href@noop {} {\bibfield  {journal} {\bibinfo  {journal} {Phys.
  Rev. Lett.}\ }\textbf {\bibinfo {volume} {99}},\ \bibinfo {pages} {200405}
  (\bibinfo {year} {2007})}\BibitemShut {NoStop}%
\bibitem [{\citenamefont {Takasu}\ and\ \citenamefont
  {Takahashi}(2009)}]{Takasu}%
  \BibitemOpen
  \bibfield  {author} {\bibinfo {author} {\bibfnamefont {Y.}~\bibnamefont
  {Takasu}}\ and\ \bibinfo {author} {\bibfnamefont {Y.}~\bibnamefont
  {Takahashi}},\ }\href@noop {} {\bibfield  {journal} {\bibinfo  {journal} {J.
  Phys. Soc. Jpn.}\ }\textbf {\bibinfo {volume} {78}},\ \bibinfo {pages}
  {012001} (\bibinfo {year} {2009})}\BibitemShut {NoStop}%
\bibitem [{2D()}]{2D}%
  \BibitemOpen
  \href@noop {} {}\bibinfo {note} {In realistic optical lattice systems, the
  dimensionality is often quasi-two-dimensional, and weakly coupled
  two-dimensional sheets of atoms are realized. If we include the third
  dimension to the analysis, the topological class changes to the
  three-dimensional weak topological phases \cite{Hasan}, which is not robust
  against disorder. However, the topological phases here are considered not to
  be affected seriously from this aspect since the optical lattices are
  considerably clean and free from disorder.}\BibitemShut {Stop}%
\bibitem [{\citenamefont {Fu}\ and\ \citenamefont {Kane}(2007)}]{FuKane}%
  \BibitemOpen
  \bibfield  {author} {\bibinfo {author} {\bibfnamefont {L.}~\bibnamefont
  {Fu}}\ and\ \bibinfo {author} {\bibfnamefont {C.~L.}\ \bibnamefont {Kane}},\
  }\href@noop {} {\bibfield  {journal} {\bibinfo  {journal} {Phys. Rev. B}\
  }\textbf {\bibinfo {volume} {76}},\ \bibinfo {pages} {045302} (\bibinfo
  {year} {2007})}\BibitemShut {NoStop}%
\bibitem [{inv()}]{inversion}%
  \BibitemOpen
  \href@noop {} {}\bibinfo {note} {Because of lack of the inversion symmetry in
  the time-dependent term, it becomes insufficient to determine the topological
  properties only from the signs of the mass term in the effective Hamiltonian.
  However, if the inversion-symmetry-breaking perturbation is small, the
  topological properties are connected equivalently to those in the completely
  inversion-symmetric case.}\BibitemShut {Stop}%
\bibitem [{not()}]{note}%
  \BibitemOpen
  \href@noop {} {}\bibinfo {note} {In the calculation of the quasi-energy
  spectrum, we deform the original triangular lattice to square lattice
  geometry, which corresponds to
  $\bm{a}_1=(1,0)a,\bm{a}_2=(0,1)a,\bm{a}_3=(-1,1)a$, and employ the open
  boundary condition for the $x$ direction and the periodic boundary condition
  for the $y$ direction. However, the bulk band does not close its band gap
  with this deformation, and therefore the topological properties are identical
  for the two lattice structures.}\BibitemShut {Stop}%
\bibitem [{\citenamefont {Schrieffer}\ and\ \citenamefont {Wolff}(1966)}]{SW}%
  \BibitemOpen
  \bibfield  {author} {\bibinfo {author} {\bibfnamefont {J.~R.}\ \bibnamefont
  {Schrieffer}}\ and\ \bibinfo {author} {\bibfnamefont {P.~A.}\ \bibnamefont
  {Wolff}},\ }\href@noop {} {\bibfield  {journal} {\bibinfo  {journal} {Phys.
  Rev.}\ }\textbf {\bibinfo {volume} {149}},\ \bibinfo {pages} {491} (\bibinfo
  {year} {1966})}\BibitemShut {NoStop}%
\bibitem [{ptb()}]{ptb_note}%
  \BibitemOpen
  \href@noop {} {}\bibinfo {note} {We remark that the expression Eq.
  (\ref{ptb_formula}) is expanded in powers of $S(t)$ but does \textit{not}
  fully contain the higher-order contribution of $r/\omega$, because the
  time-dependent term we neglected in this approximation can contribute in
  $O((r/\omega)^2)$ to the effective Hamiltonian. This problem can be overcome
  by repeating the procedure of the perturbation theory including the neglected
  time-dependent term as a new perturbation.}\BibitemShut {Stop}%
\bibitem [{\citenamefont {Gerbier}\ and\ \citenamefont
  {Dalibard}(2010)}]{Gerbier}%
  \BibitemOpen
  \bibfield  {author} {\bibinfo {author} {\bibfnamefont {F.}~\bibnamefont
  {Gerbier}}\ and\ \bibinfo {author} {\bibfnamefont {J.}~\bibnamefont
  {Dalibard}},\ }\href@noop {} {\bibfield  {journal} {\bibinfo  {journal} {New
  J. Phys.}\ }\textbf {\bibinfo {volume} {12}},\ \bibinfo {pages} {033007}
  (\bibinfo {year} {2010})}\BibitemShut {NoStop}%
\bibitem [{\citenamefont {Buchhold}\ \emph {et~al.}(2012)\citenamefont
  {Buchhold}, \citenamefont {Cocks},\ and\ \citenamefont
  {Hofstetter}}]{Buchhold}%
  \BibitemOpen
  \bibfield  {author} {\bibinfo {author} {\bibfnamefont {M.}~\bibnamefont
  {Buchhold}}, \bibinfo {author} {\bibfnamefont {D.}~\bibnamefont {Cocks}}, \
  and\ \bibinfo {author} {\bibfnamefont {W.}~\bibnamefont {Hofstetter}},\
  }\href@noop {} {\bibfield  {journal} {\bibinfo  {journal} {Phys. Rev. A}\
  }\textbf {\bibinfo {volume} {85}},\ \bibinfo {pages} {063614} (\bibinfo
  {year} {2012})}\BibitemShut {NoStop}%
\bibitem [{\citenamefont {Wang}\ \emph {et~al.}(2010)\citenamefont {Wang},
  \citenamefont {Qi},\ and\ \citenamefont {Zhang}}]{Wang1}%
  \BibitemOpen
  \bibfield  {author} {\bibinfo {author} {\bibfnamefont {Z.}~\bibnamefont
  {Wang}}, \bibinfo {author} {\bibfnamefont {X.~L.}\ \bibnamefont {Qi}}, \ and\
  \bibinfo {author} {\bibfnamefont {S.~C.}\ \bibnamefont {Zhang}},\ }\href@noop
  {} {\bibfield  {journal} {\bibinfo  {journal} {Phys. Rev. Lett.}\ }\textbf
  {\bibinfo {volume} {105}},\ \bibinfo {pages} {256803} (\bibinfo {year}
  {2010})}\BibitemShut {NoStop}%
\bibitem [{\citenamefont {Wang}\ and\ \citenamefont {Zhang}(2012)}]{Wang2}%
  \BibitemOpen
  \bibfield  {author} {\bibinfo {author} {\bibfnamefont {Z.}~\bibnamefont
  {Wang}}\ and\ \bibinfo {author} {\bibfnamefont {S.~C.}\ \bibnamefont
  {Zhang}},\ }\href@noop {} {\bibfield  {journal} {\bibinfo  {journal} {Phys.
  Rev. X}\ }\textbf {\bibinfo {volume} {2}},\ \bibinfo {pages} {031008}
  (\bibinfo {year} {2012})}\BibitemShut {NoStop}%
\bibitem [{\citenamefont {Wang}\ and\ \citenamefont {Yan}(2013)}]{Wang3}%
  \BibitemOpen
  \bibfield  {author} {\bibinfo {author} {\bibfnamefont {Z.}~\bibnamefont
  {Wang}}\ and\ \bibinfo {author} {\bibfnamefont {B.}~\bibnamefont {Yan}},\
  }\href@noop {} {\bibfield  {journal} {\bibinfo  {journal} {J. Phys. Condens.
  Matter}\ }\textbf {\bibinfo {volume} {25}},\ \bibinfo {pages} {155601}
  (\bibinfo {year} {2013})}\BibitemShut {NoStop}%
\end{thebibliography}%

\end{document}